\documentclass[preprint]{elsarticle}
\usepackage{amssymb}
\usepackage{amsmath,amsthm}
\usepackage{mathrsfs}
\usepackage{slashbox}
\usepackage{amstext}
\usepackage{mathptmx}
\usepackage{bm}
\usepackage{booktabs}
\usepackage{multirow, mathtools}
\usepackage{lineno,hyperref}
\usepackage{color}
\usepackage{flafter}
\usepackage{tabularx}
\usepackage[hang]{subfigure}
\usepackage{algorithm}
\usepackage{algpseudocode}
\usepackage{algorithmicx}
\algnewcommand\algorithmicswitch{\textbf{switch}}
\algnewcommand\algorithmiccase{\textbf{case}}
\algdef{SE}[SWITCH]{Switch}{EndSwitch}[1]{\algorithmicswitch\ #1}{\algorithmicend\ \algorithmicswitch}%
\algdef{SE}[CASE]{Case}{EndCase}[1]{\algorithmiccase\ #1}{\algorithmicend\ \algorithmiccase}%
\algtext*{EndCase}%

\usepackage{ulem}
\modulolinenumbers[5]
\usepackage{pgfplots}
\pgfplotsset{compat=newest}
\usetikzlibrary{plotmarks}
\usetikzlibrary{arrows.meta}
\usepgfplotslibrary{patchplots}
\usepackage{grffile}
\newlength\fwidth

\journal{a journal}

\graphicspath{{figures/}}

\begin{document}

\begin{frontmatter}
\title{Efficient single-grid and multi-grid solvers for real-space orbital-free density functional theory}

\author[mymainaddress]{Ling-Ze Bu\corref{mycorrespondingauthor}}
\cortext[mycorrespondingauthor]{Corresponding author}
\ead{17b933010@stu.hit.edu.cn}
\author[mymainaddress,mysecondaryaddress,mythirdaddress]{Wei Wang}
\ead{wwang@hit.edu.cn}
\address[mymainaddress]{School of Civil Engineering, Harbin Institute of Technology, Harbin 150090, China}
\address[mysecondaryaddress]{Key Lab of Structures Dynamic Behavior and Control of the Ministry of Education, Harbin Institute of Technology, Harbin 150090, China}
\address[mythirdaddress]{Key Lab of Smart Prevention and Mitigation of Civil Engineering Disasters of the Ministry of Industry and Information Technology, Harbin Institute of Technology, Harbin 150090, China}

\begin{abstract}
	To improve the computational efficiencies of the real-space orbital-free density functional theory, this work develops a new single-grid solver by directly providing the closed-form solution to the inner iteration and using an improved bisection method to accelerate the line search process in the outer iteration, and extended the single-grid solver to a multi-grid solver. Numerical examples show that the proposed single-grid solver can improve the computational efficiencies by two orders of magnitude comparing with the methods in the literature and the multi-grid solver can improve the computational efficiencies even once for the cases where high-resolution electron densities are needed.
\end{abstract}


\end{frontmatter}


\section{Introduction}\label{Intro}
Density functional theory (DFT) regards the three-dimensional ground state electron density rather than the $ 3N_{\mathrm{e}} $-dimensional ($ N_{\mathrm{e}} $ denotes the number of electrons in the system) wave functions as the independent variable, thus has a great advantage over the wave function-based methods in quantum chemistry in terms of computational efficiencies, hence becomes a workhorse in computational chemistry. In traditional DFT, the kinetic energy is expressed as a functional of $ N_{\mathrm{e}} $ orthogonal orbitals, the orbitals are solutions of the Kohn-Sham equation and are combined into the ground state electron density. However, this kind of methods not only have low computational efficiencies, bu also may not converge. Although the direct minimization methods are free of self-consistent field iterations, the computational efficiencies cannot be substantially enhanced under the framework of the traditional DFT. In fact, the idea of expressing the kinetic energy as a direct functional of the ground state electron density had been proposed in 1927\cite{Thomas1927}, nonetheless, orbital-free DFT (OFDFT) has become a mainstream method since the beginning of the 21st century. Numerical methods include plane-wave ones\cite{Ho2008,Hung2009,Chen2016} and real-space ones. The former utilize fast Fourier transform to compute the kinetic energy, thus have high computational efficiencies, nonetheless, these methods are only suitable for periodic systems. The later are suitable for non-periodic systems thus has been attracting more attention in recent years.

The real-space methods can be classified into finite element methods and finite difference methods according to discretization schemes, or can be classified into direct methods and transform methods according to the treatment of the electrostatic terms. The basic idea of the transform methods is to rewrite the electrostatic terms as a variational problem with respect to the total electrostatic potential, thus can avoid the non-local computations including the convolution in the Hartree term and the Ewald summation in the ion-ion interaction, making the overall computational complexity reduces to linear scaling with respect to the number of ions. Based on the second-order finite difference discretization and the direct method, Jiang et al. \cite{Jiang2004} firstly used the conjugate gradient method to compute the ground state electron density. 
Garc{\i}a-Cervera et al.\cite{GarciaCervera2007} used the truncated Newton method to compute the ground state electron density. Mi et al. \cite{Mi2016} developed a software package ATLAS for computing the ground state electron density of periodic systems. Shao et al. \cite{Shao2021} proposed an improved self-consistent field iteration method for computing the ground state electron density. Gavini et al. \cite{Gavini2007} firstly proposed the general finite element framework for non-periodic systems and firstly proposed the transform method, and proved some theoretical properties of the finite element method. Motamarri et al. \cite{Motamarri2013} extended this method by using higher-order finite elements, leading to improvements of computational efficiencies by two orders of magnitude, and proved the error boundaries theoretically. Das et al. \cite{Das2015} extended the higher-order finite element method by using the Wang-Govind-Carter type non-local kinetic energy, and rewrite the saddle-point problem as a fixed-point problem, the later was solved with a self-consistent field iteration method. Suryanarayana et al. \cite{Suryanarayana2014} carried on the transform method and proposed a higher-order finite difference method for general non-periodic systems, and rewrote the constrained optimization problem as a unconstrained optimization problem by using the augmented Lagrange method, and the problem was solved with the nonlinear conjugate gradient method. Ghosh et al.\cite{Ghosh2016} also carried on the transform method, and proposed the higher-order finite difference method for periodic systems, and computed the ground state electron density by using a fixed-point iteration method and a parallel computation strategy. 

Previous works mainly focused on numerical discretization schemes and used ready-made solvers, and did not specially studied the solvers, leading to low computational efficiencies. New solvers will help enhancing the computational efficiencies. Developing new solvers should overcome two difficulties: first, the inner solver for Poission equation should directly provide a closed-from solution without iteration; second, the computational efficiencies of the outer nonlinear conjugate gradient method should be improved by improving the line search efficiencies and using preconditioning schemes. In this context, this work will develop a new single-grid solver from the following two aspects: first, discretize the inner Poisson equation into a Sylvester tensor equation by using a higher-order finite difference scheme, and derive the closed-form solution with the generalized eigenvalue decomposition in three directions; second, proposing an improved bisection method to improve the line search efficiencies in the outer nonlinear conjugate gradient iterations. Then, the single-grid solver will be extended into a multi-grid solver. Finally, the computational efficiencies of the proposed solvers over that of the literature will be illustrated by numerical examples.  

\section{Overview of orbital-free density functional theory}
According to OFDFT\cite{Gavini2007}, the ground state energy of an $ M $-atom system is expressed as Eq.(\ref{E}),
\begin{equation}\label{E}
	E(\rho, \bm{R})=T_{\mathrm{s}}(\rho)+E_{\mathrm{xc}}(\rho)+E_{\mathrm{H}}(\rho)+E_{\mathrm{ext}}(\rho, \bm{R})+E_{\mathrm{zz}}(\bm{R}),
\end{equation}
\begin{tabularx}{\textwidth}{@{}l@{\quad}r@{---}X@{}}
	where& $\rho$ &$\rho(\bm{x})$, representing the electron density at spatial coordinate $\bm{x}$;\\
	&  $\bm{R}$&$\left\{\bm{R}_{1}, \ldots, \bm{R}_{M}\right\}$, representing the set of the nucleus positions;\\
	&  $E$ &the total energy functional;\\
	&  $T_{\mathrm{s}}$ & the electronic kinetic energy;\\
	&  $E_{\mathrm{xc}}$ & the exchange-correlation functional;\\
	&  $E_{\mathrm{H}}$ & electrostatic interaction energy between electrons, also named as Hartree energy;\\
	&  $E_{\mathrm{ext}}$ & electrostatic interaction energy between electrons and external electronic field; \\
	&  $E_{\mathrm{zz}}$ &repulsion energy between nuclei.
\end{tabularx}

$T_{\mathrm{s}}$is usually selected from Thomas-Fermi-Weizsacker functional family, as shown in Eq.(\ref{Ts}),
\begin{equation}\label{Ts}
	T_{\mathrm{s}}(\rho)=C_{\mathrm{F}} \int_{\Omega} \rho^{\frac{5}{3}}(\bm{x}) \mathrm{d} \bm{x}+\frac{\lambda}{8} \int_{\Omega} \frac{|\nabla \rho(\bm{x})|^{2}}{\rho(\bm{x})} \mathrm{d} \bm{x},
\end{equation}
\begin{tabularx}{\textwidth}{@{}l@{\quad}r@{---}X@{}}
	where& $C_{\mathrm{F}}$ &constant, equals to $\frac{3}{10}(3\pi^2)^{\frac{2}{3}}$;\\
	&  $\lambda$&parameter, selected as 0.2 in this paper;\\
	&  $\Omega$   &the support of $\rho$.
\end{tabularx}

$E_{\mathrm{xc}}$ is selected as the local density approximation, as shown in Eq.(\ref{Exc}),
\begin{equation}\label{Exc}
	E_{\mathrm{xc}}(\rho)=\int_{\Omega} \varepsilon_{\mathrm{xc}}(\rho(\bm{x})) \rho(\bm{x}) \mathrm{d} \bm{x},
\end{equation}
where $\varepsilon_{\mathrm{xc}}=\varepsilon_{\mathrm{x}}+\varepsilon_{\mathrm{c}}$ is the exchange-correlation energy per electron, as shown in Eq.(\ref{exec}),
\begin{subequations}\label{exec}
	\begin{alignat}{1}
		\varepsilon_{\mathrm{x}}(\rho)=&-\frac{3}{4}\left(\frac{3}{\pi}\right)^{\frac{1}{3}} \rho^{\frac{1}{3}},\\
		\varepsilon_{\mathrm{c}}(\rho)=&\begin{cases}
			\dfrac{\gamma}{1+\beta_{1} \sqrt{r_{\mathrm{s}}}+\beta_{2} r_{\mathrm{s}}}, & r_{\mathrm{s}} \geqslant 1 \\
			A \log r_{\mathrm{s}}+B+C r_{\mathrm{s}} \log r_{\mathrm{s}}+D r_{\mathrm{s}}, & r_{\mathrm{s}}<1
		\end{cases},
	\end{alignat}
\end{subequations}
where $r_{\mathrm{s}}=(3/4 \pi \rho)^{1/3}$, $\gamma=-0.1423$, $\beta_{1}=1.0529$, $\beta_{2}=0.3334$, $A=0.0311$, $B=-0.048$, $C=0.002$, $D=-0.0116$.

The last three terms of the energy functional Eq.(\ref{E}) are the electrostatic interaction terms:
\begin{equation}\label{EH}
	E_{\mathrm{H}}(\rho)=\dfrac{1}{2}  \iint_{\Omega\times\Omega} \frac{\rho(\bm{x}) \rho\left(\bm{x}^{\prime}\right)}{\left|\bm{x}-\bm{x}^{\prime}\right|} \mathrm{d} \bm{x} \mathrm{d} \bm{x}^{\prime},
\end{equation}
\begin{equation}\label{Eext}
	E_{\mathrm{ext}}(\rho, \bm{R})=\int_{\Omega} \rho(\bm{x}) \left( \sum_{J=1}^{M} V_J(\bm{x},\bm{R}_J)\right) \mathrm{d} \bm{x},
\end{equation}
\begin{equation}\label{Ezz}
	E_{\mathrm{zz}}(\bm{R})=\dfrac{1}{2} \sum_{I=1}^{M} \sum_{J=1 \atop J \neq I}^{M} \frac{Z_{I} Z_{J}}{\left|\bm{R}_{I}-\bm{R}_{J}\right|}.
\end{equation}
In Eq.(\ref{Eext}), $V_J(\bm{x},\bm{R}_J)$ represents the electric field generated by the nucleus at $\bm{R}_{J}$. In Eq.(\ref{Ezz}), $Z_I$ represents the charge of nucleus at $\bm{R}_I$. According to the convention of electronic structure computation\cite{Suryanarayana2010}, nuclear charges are negative and electron charges are positive.

In all-electron computation, nuclei are regarded as point charges, thus $V_J(\bm{x},\bm{R}_J)$ is derived from Coulomb potential . However, due to the singularity of the potential at the nucleus, the potential around the nucleus need many basis functions to be accurately represented. Meanwhile, core electrons are chemically inactive, and orbital-free kinetic energy functionals are not accurate for highly non-smooth electron densities. In this context, a usual way is to consider only the covalent electrons, then $V_J(\bm{x},\bm{R}_J)$ is an effective potential named as pseudopotential. $Z_I$ represents the ion charge at $\bm{R}_I$.  The orbital-free feature of the kinetic functional requires the pseudopotential to be local, i.e. $V_J(\bm{x},\bm{R}_J)=V_\mathrm{a}(|\bm{x}-\bm{R}_J|)$. The ion charge density is expressed as Eq.(\ref{b}),
\begin{equation}\label{b}
	b(\bm{x}, \bm{R})=\sum_{J=1}^{M} b_{J}(\bm{x}, \bm{R}_{J}), \quad b_{J}(\bm{x}, \bm{R}_{J})=-\dfrac{1}{4 \pi} \nabla^{2} V_{J}(\bm{x}, \bm{R}_{J}).
\end{equation}
The local pseudopotential and the Coulomb potential are the same in the space outside the cut-off radius $r_{\mathrm{c}}$, thus $ b_{J}(\bm{x}, \bm{R}_{J}) $ has a compact support with center $ \bm{R}_J $ and radius $ r_{\mathrm{c} }$, and satisfy Eq.(\ref{bJ}),
\begin{equation}\label{bJ}
	\int_{\mathbb{R}^{3}} b_{J}\left(\bm{x}, \bm{R}_{J}\right) \mathrm{d} \bm{x}=Z_{J}.
\end{equation}	

The computational cost of directly computing Eqs.(\ref{EH}) to (\ref{Ezz}) will be $ O(M^2) $. To reduce the computational cost, the total electrostatic energy can be rewritten as the linear-scaling form in Eq.(\ref{Ephi}),
\begin{equation}\label{Ephi}
	\begin{aligned}
		E_{\mathrm{H}}(\rho)+E_{\mathrm{ext}}(\rho, \bm{R})+E_{\mathrm{zz}}(\bm{R}) &=-\inf _{\phi \in Y}\left\{\frac{1}{8 \pi} \int_{\Omega}|\nabla \phi(\bm{x})|^{2} \mathrm{d} \bm{x}-\int_{\Omega}\left(\rho(\bm{x})+b(\bm{x}, \bm{R})\right) \phi(\bm{x}) \mathrm{d} \bm{x}\right\} \\
		&-\dfrac{1}{2} \sum_{J=1}^{M} \int_{\Omega} b_{J}(\bm{x}, \bm{R}_{J}) V_{J}(\bm{x}, \bm{R}_{J}) \mathrm{d} \bm{x}+E_{\mathrm{c}}(\bm{R}),
	\end{aligned}
\end{equation}
\begin{tabularx}{\textwidth}{@{}l@{\quad}r@{---}X@{}}
	where& $\phi$ &the total electrostatic potential generated by the covalent electrons and the ions;\\
	&  $Y$& an admissible function space;\\
	&  $E_{\mathrm{c}}(\bm{R})$ &the correction term for the cases when the supports of $b_{J}$ overlap with each other.
\end{tabularx}

By taking the variation of Eq.(\ref{Ephi}), we derive that $ \phi $ is the solution of the Poisson equation in Eq.(\ref{phi}),
\begin{equation}\label{phi}
	-\dfrac{1}{4\pi}\nabla^2\phi = \rho(\bm{x})+b(\bm{x}, \bm{R})
\end{equation}
When the supports of $b_{J}$ do not overlap, we derive Eq.(\ref{Ebb}),
\begin{equation}\label{Ebb}
	E_{\mathrm{zz}}(\bm{R})=\dfrac{1}{2} \iint_{\Omega\times\Omega} \frac{b(\bm{x}, \bm{R}) b(\bm{x}^{\prime}, \bm{R})}{\left|\bm{x}-\bm{x}^{\prime}\right|} \mathrm{d} \bm{x} \mathrm{d} \bm{x}^{\prime}-\frac{1}{2} \sum_{J=1}^{M} \int_{\Omega} b_{J}(\bm{x}, \bm{R}_{J}) V_{J}(\bm{x}, \bm{R}_{J}) \mathrm{d} \bm{x},
\end{equation}
otherwise, we can choose $\tilde{b}_{J}\left(\bm{x}, \bm{R}_{J}\right)$ with disjoint supports and satisfy Eq.(\ref{bJ2}),
\begin{equation}\label{bJ2}
	\int_{\mathbb{R}^{3}} \tilde{b}_{J}(\bm{x}, \bm{R}_{J}) \mathrm{d} \bm{x}=Z_{J}.
\end{equation}

Let
\begin{equation}
	\tilde{b}(\bm{x}, \bm{R})=\sum_{J=1}^{M} \tilde{b}_{J}(\bm{x}, \bm{R}_{J}), \quad \tilde{b}_{J}(\bm{x}, \bm{R}_{J})=\frac{-1}{4 \pi} \nabla^{2} \tilde{V}_{J}(\bm{x}, \bm{R}_{J}).
\end{equation}
then we derive Eq.(\ref{Ec}),
\begin{equation}\label{Ec}
	\begin{aligned}
		E_{\mathrm{c}}(\bm{R})=& \left(\dfrac{1}{2} \iint_{\Omega\times\Omega} \frac{\tilde{b}(\bm{x}, \bm{R}) \tilde{b}(\bm{x}^{\prime}, \bm{R})}{\left|\bm{x}-\bm{x}^{\prime}\right|} \mathrm{d} \bm{x} \mathrm{d} \bm{x}^{\prime}-\dfrac{1}{2} \sum_{J=1}^{M} \int_{\Omega} \tilde{b}_{J}(\bm{x}, \bm{R}_{J}) \tilde{V}_{J}(\bm{x}, \bm{R}_{J}) \mathrm{d} \bm{x} \right)\\
		&-\left(\dfrac{1}{2} \iint_{\Omega\times\Omega} \frac{b(\bm{x}, \bm{R}) b(\bm{x}^{\prime}, \bm{R})}{\left|\bm{x}-\bm{x}^{\prime}\right|} \mathrm{d} \bm{x} \mathrm{d} \bm{x}^{\prime}-\dfrac{1}{2} \sum_{J=1}^{M} \int_{\Omega} b_{J}(\bm{x}, \bm{R}_{J}) V_{J}(\bm{x}, \bm{R}_{J}) \mathrm{d} \bm{x}\right).
	\end{aligned}
\end{equation}
Eq.(\ref{Ec}) can be written as the equivalent linear-scaling form in Eq.(\ref{Ec2}),
\begin{equation}\label{Ec2}
	\begin{aligned}
		E_{\mathrm{c}}(\bm{R})=& \dfrac{1}{2} \int(\tilde{b}(\bm{x}, \bm{R})+b(\bm{x}, \bm{R})) \phi_{\mathrm{c}}(\bm{x}, \bm{R}) \mathrm{d} \bm{x}+\dfrac{1}{2} \sum_{J=1}^{M} \int b_{J}(\bm{x}, \bm{R}_{J}) \phi_{J}(\bm{x}, \bm{R}_{J}) \mathrm{d} \bm{x} \\
		&-\dfrac{1}{2} \sum_{J=1}^{M} \int \tilde{b}_{J}(\bm{x}, \bm{R}_{J}) \tilde{\phi}_{J}(\bm{x}, \bm{R}_{J}) \mathrm{d} \bm{x},
	\end{aligned}
\end{equation}
where $ \phi_{\mathrm{c}}(\bm{x}, \bm{R}) $ is the solution of the following Poisson equation:
\begin{equation}\label{phic}
	-\dfrac{1}{4 \pi} \nabla^{2} \phi_{\mathrm{c}}(\bm{x}, \bm{R})=\tilde{b}(\bm{x}, \bm{R})-b(\bm{x}, \bm{R}).
\end{equation}

$ \rho $ should satisfy the constraint in Eq.(\ref{rhoc})
\begin{equation}\label{rhoc}
	\rho\geqslant 0,\quad \int_{\Omega} \rho(\bm{x}) \mathrm{d}\bm{x} = N_{\mathrm{e}},
\end{equation}
\begin{tabularx}{\textwidth}{@{}l@{\quad}r@{---}X@{}}
	where & $N_{\mathrm{e}}$ & the total number of covalent electrons.
\end{tabularx}
Letting $ \rho=u^2 $ naturally satisfies the first constraint, thus for a given $ \bm{R} $, $ u $ is the solution of Eq.(\ref{E0}),
\begin{equation}\label{E0}
	E(\bm{R})= \inf _{u \in \mathcal{X}} E(u, \bm{R}), \quad \mathcal{X}=\left\{u: u \in X,  \int_{\Omega} u^2(\bm{x}) \mathrm{d}\bm{x} = N_{\mathrm{e}} \right\},
\end{equation}
\begin{tabularx}{\textwidth}{@{}l@{\quad}r@{---}X@{}}
	where & $ X $ & an admissible function space,
\end{tabularx}
\begin{equation}\label{EuR}
	\begin{aligned}
		E(u, \bm{R})= &\dfrac{\lambda}{2} \int_{\Omega} |\nabla u(\bm{x})|^{2} \mathrm{d} \bm{x}+C_{\mathrm{F}} \int_{\Omega} u^{\frac{10}{3}}(\bm{x}) \mathrm{d} \bm{x}+\int_{\Omega} \varepsilon_{\mathrm{xc}}(u^2(\bm{x})) u^2(\bm{x}) \mathrm{d} \bm{x}\\
		&-\inf _{\phi \in Y}\left\{\frac{1}{8\pi} \int_{\Omega}|\nabla \phi(\bm{x})|^{2} \mathrm{d} \bm{x}-\int_{\Omega}\left(u^2(\bm{x})+b(\bm{x}, \bm{R})\right) \phi(\bm{x}) \mathrm{d} \bm{x}\right\}\\
		& -\dfrac{1}{2} \sum_{J=1}^{M} \int_{\Omega} b_{J}(\bm{x}, \bm{R}_{J}) V_{J}(\bm{x}, \bm{R}_{J}) \mathrm{d} \bm{x}+E_{\mathrm{c}}(\bm{R}).
	\end{aligned}
\end{equation}

\section{The single-grid solver}
According to the previous section, electronic structure computation is to solve the optimization problem Eq.(\ref{E0}). Eq.(\ref{E0}) can be transformed to an equivalent unconstrained optimization problem as in Eq.(\ref{E02}),
\begin{equation}\label{E02}
	E(\bm{R})= \inf _{u \in X} E\left(u\sqrt{\dfrac{N_{\mathrm{e}}}{\lVert u \rVert_2}}, \bm{R}\right).
\end{equation}
This section proposes a single-grid method for solving $ u $.

The electronic structure is described using an Euler grid. The computational domain $ \Omega $ is meshed with $ m_1\times m_2\times m_3 $ eight-nodes hexahedral elements ,each element has size $ h $ in each direction. Since $ |u| $ rapidly decreases to zero for sufficiently large $ \Omega $, homogeneous Dirichlet boundary condition is utilized, thus
\begin{equation}\label{uijk}
	u(\bm{x}) = \sum_{i=1}^{n_1} \sum_{j=1}^{n_2}\sum_{k=1}^{n_3} \bm{u}(i,j,k) N_i(x_1)N_j(x_2)N_k(x_3),
\end{equation}
\begin{tabularx}{\textwidth}{@{}l@{\quad}r@{---}X@{}}
	where& $ n_i $ & equals to $ m_i-1 $, denotes the number of degree of freedom in the $ i $th direction,\\
	& $ \bm{u}(i,j,k)$ & the value of $ u $ at node $ (i,j,k) $,\\
	& $ N_i $ & the single-direction basis function of node $ i $.
\end{tabularx}

According to Eq.(\ref{EuR}), before formally solving $ u $, $ V_J $and $ b_J$ should be determined. This work use Goodwin-Needs-Heine pseudopotential shown in Eq.(\ref{GNH}),
\begin{equation}\label{GNH}
	V_{\mathrm{a}}(|\bm{x}|)=-\dfrac{2}{\pi} \int_{0}^{\infty} \frac{\sin (|\bm{x}| r)}{|\bm{x}| r}\left((Z-AR) \cos (Rr)+A \frac{\sin (Rr)}{r}\right) \exp{-(\frac{r}{r_{\mathrm{c}}} )^{6}} \mathrm{d}r,
\end{equation}
\begin{tabularx}{\textwidth}{@{}l@{\quad}r@{---}X@{}}
	where & $ Z$ & the number of covalent electrons of an atom,\\
	& $ r_{\mathrm{c}}$ & cut-off radius,\\
	& $ A,R $ & parameters.
\end{tabularx}
Let $ r= |\bm{x}|$, then
\begin{equation}\label{br}
	b_{\mathrm{a}}(\bm{x})=-\dfrac{1}{4 \pi} \nabla^{2} V_{\mathrm{a}}(\bm{x}) \Rightarrow b_{\mathrm{a}}(r)=-\dfrac{1}{4 \pi}\dfrac{1}{r^2} \dfrac{\mathrm{d}}{\mathrm{d}r}\left(r^2 \dfrac{\mathrm{d}}{\mathrm{d}r}\right) V_{\mathrm{a}}(r).
\end{equation}
The computation of continuous functions in the above equations can be done with Chebfun toolbox\cite{Battles2004}. Then, by using $ V_J(\bm{x},\bm{R}_J)=V_\mathrm{a}(|\bm{x}-\bm{R}_J|) $ and $ b_J(\bm{x},\bm{R}_J)=b_\mathrm{a}(|\bm{x}-\bm{R}_J|) $, we derive $ V_J $ and $ b_J $.

After obtaining $ V_J $and $ b_J $, this work use the nonlinear conjugate gradient method to solver the optimization problem in Eq.(\ref{E02}). From Eq.(\ref{uijk}), we derive
\begin{equation}\label{u2}
	\begin{split}
		\lVert u \rVert& = \int_{\Omega} u^2(\bm{x}) \mathrm{d}\bm{x}\\
		&=\langle \bm{u}, \bm{M}\bullet\bm{u}\rangle
	\end{split}
\end{equation}
\begin{tabularx}{\textwidth}{@{}l@{\quad}r@{---}X@{}}
	where & $ \langle\quad\rangle$ & inner product: $ \langle \bm{u},\bm{v} \rangle = \mathrm{vec}(\bm{u})^{\mathrm{T}} \mathrm{vec}(\bm{v})$, \\
	& $ \bm{M} $ & equals to $ \left[\mathrm{tridiag}\left(\frac{1}{6}, \frac{2}{3},\frac{1}{6}\right), \mathrm{tridiag}\left(\frac{1}{6}, \frac{2}{3},\frac{1}{6}\right),\mathrm{tridiag}\left(\frac{1}{6}, \frac{2}{3},\frac{1}{6}\right); h^3\right] $, is a third-order Tucker-type matrix, \\
	& $ \bm{M}\bullet\bm{u} $ & the product of vector $ \bm{M} $ and vector $ \bm{u} $.
\end{tabularx}
Any matrix $ \bm{T} $ with form as Eq.(\ref{Txx})
\begin{equation}\label{Txx}
	\bm{T} = \sum_{i=1}^{r_1} \sum_{j=1}^{r_2} \sum_{k=1}^{r_3} \bm{C}(i,j,k) \bm{T}_1(:,:,i)\otimes\bm{T}_2(:,:,j)\otimes\bm{T}_3(:,:,k)
\end{equation}
can be expressed as Tucker form $ \bm{T}=[\bm{T}_1,\bm{T}_2, \bm{T}_3; \bm{C}] $. $ \bm{T}\bullet\bm{u} $ is defined as Eq.(\ref{Tu})
\begin{equation}\label{Tu}
	\bm{T}\bullet\bm{u} \triangleq \sum_{i=1}^{r_1} \sum_{j=1}^{r_2} \sum_{k=1}^{r_3} \bm{C}(i,j,k) \bm{u} \times_1 \bm{T}_1(:,:,i)\times_2 \bm{T}_2(:,:,j)\times_3 \bm{T}_3(:,:,k).
\end{equation}

For given $ \bm{R} $, let $ \bm{u}_{\mathrm{n}} = \bm{u}\sqrt{N_{\mathrm{e}}/\langle\bm{u},\bm{M}\bullet\bm{u}\rangle}$, we define
\begin{equation}\label{Fu}
	F(\bm{u}) = E(\bm{u}_{\mathrm{n}},\bm{R})
\end{equation}
where 
\begin{equation}\label{Eun}
	E(\bm{u}_{\mathrm{n}},\bm{R}) = \dfrac{\lambda}{2}\langle \bm{u}_{\mathrm{n}}, \bm{A}\bullet\bm{u}_{\mathrm{n}}\rangle + \left(\sum f(\bm{u}_{\mathrm{n}})\right)\cdot h^3 + \dfrac{1}{2}\langle \bm{M}\bullet\bm{u}_{\mathrm{n}}.^2 + \bm{b}, \bm{\phi}\rangle
\end{equation}
where 
\begin{equation}\label{ABQC}
	\begin{split}
		\bm{A}&=\left[\bm{B}_1,\bm{B}_2,\bm{B}_3;\bm{C}\right],\\
		\bm{B}_i&=[\bm{P}_i| \bm{Q}_i],\ i=1,2,3\\
		\bm{Q}_i&=\mathrm{tridiag}\left(\frac{1}{6},\frac{2}{3},\frac{1}{6}\right)_{n_i},
		\bm{C}=\left[\begin{array}{ll|ll}
			0 & 0 & 0 & 1 \\
			0 & 1 & 1 & 0
		\end{array}\right]\cdot h,
	\end{split}
\end{equation}
\begin{equation}
	f(\bm{u}_{\mathrm{n}})(i,j,k)=C_{\mathrm{F}} \bm{u}^{\frac{10}{3}}_{\mathrm{n}}(i,j,k) +\varepsilon_{\mathrm{xc}}(\bm{u}^2_{\mathrm{n}}(i,j,k)) \cdot \bm{u}^2_{\mathrm{n}}(i,j,k),
\end{equation}
$ \sum $ is the summation of all the elements in the vector, $ \bm{u}_{\mathrm{n}}.^2 $ is the square of all elements in $ \bm{u}_{\mathrm{n}} $, $ \bm{\phi}$ is the solution of the following discretized Poission equation with homogeneous Dirichlet boundary condition 
\begin{equation}\label{Poi}
	\dfrac{1}{4\pi}\bm{A}\bullet\bm{\phi} = \bm{M}\bullet\bm{u}_{\mathrm{n}}.^2 + \bm{b},
\end{equation}
where
\begin{equation}\label{bijk}
	\begin{split}
		\bm{b}(i,j,k)=&\int_{\Omega} b(\bm{x},\bm{R})N_i(x_1)N_j(x_2)N_k(x_3)\mathrm{d}\bm{x},\\
		=&\left(\dfrac{h}{2}\right)^3 (b((i-0.5)h,(j-0.5)h,(k-0.5)h;\bm{R})+b((i+0.5)h,(j-0.5)h,(k-0.5)h;\bm{R})+\\
		&b((i-0.5)h,(j+0.5)h,(k-0.5)h;\bm{R})+b((i-0.5)h,(j-0.5)h,(k+0.5)h;\bm{R})+\\
		&b((i+0.5)h,(j+0.5)h,(k-0.5)h;\bm{R})+b((i+0.5)h,(j-0.5)h,(k+0.5)h;\bm{R})+\\
		&b((i-0.5)h,(j+0.5)h,(k+0.5)h;\bm{R})+b((i+0.5)h,(j+0.5)h,(k+0.5)h);\bm{R}),
	\end{split}
\end{equation}
$ \bm{P}_i $ is a finite difference discretization of the negative Laplace operator. According to \cite{Suryanarayana2014}, the $ n $th order finite difference discretization of $ -\nabla^2\phi $ is 
\begin{equation}
	\left.-\nabla^{2} \phi\right|^{(i, j, k)} \approx \sum_{p=0}^{n} w_{p}(\phi^{(i+p, j, k)}+\phi^{(i-p, j, k)}+\phi^{(i, j+p, k)}+\phi^{(i, j-p, k)}+\phi^{(i, j, k+p)}+\phi^{(i, j, k-p)})
\end{equation}
\begin{tabularx}{\textwidth}{@{}l@{\quad}r@{---}X@{}}
	where & $ \phi^{(i,j,k)}$ & the value of $ \phi $ at $ (i,j,k) $,
\end{tabularx}
$ w_p $ is determined by Eq.(\ref{wp}).
\begin{equation}\label{wp}
	\begin{aligned}
		w_{0} &=\frac{1}{h^{2}} \sum_{q=1}^{n} \frac{1}{q^{2}}, \\
		w_{p} &=\frac{2(-1)^{p}}{h^{2} p^{2}} \frac{(n !)^{2}}{(n-p)!(n+p)!}, \quad p=1,2, \ldots, n .
	\end{aligned}
\end{equation}

Let $ \bm{t}=\bm{M}\bullet\bm{u}_{\mathrm{n}} + \bm{b} $, then Eq.(\ref{Poi}) can be rewritten as Eq.(\ref{Poi2}),
\begin{equation}\label{Poi2}
	\dfrac{1}{4\pi}(\bm{P}_1\otimes\bm{Q}_2\otimes\bm{Q}_3+\bm{Q}_1\otimes\bm{P}_2\otimes\bm{Q}_3+\bm{Q}_1\otimes\bm{Q}_2\otimes\bm{P}_3)\bm{\phi}=\bm{t}.
\end{equation}

After determining $ \bm{P}_i\ (i=1,2,3) $ with Eq.(\ref{wp}), for each $ i $, we proceed generalized singular value decomposition as Eq.(\ref{gevd}),
\begin{equation}\label{gevd}
	\bm{P}_i \bm{V}_i = \bm{Q}_i \bm{V}_i \bm{\Lambda}_i,
\end{equation}
Eq.(\ref{QP}) is derived from (\ref{gevd})
\begin{equation}\label{QP}
	\bm{Q}_i = (\bm{V}_i^{-1})^{\mathrm{T}} \bm{V}_i^{-1},\ \bm{P}_i = (\bm{V}_i^{-1})^{\mathrm{T}}\bm{\Lambda} \bm{V}_i^{-1},
\end{equation}
substitute Eq.(\ref{QP}) into Eq.(\ref{Poi2}), we get
\begin{equation}\label{phisol}
	\begin{aligned}
		\bm{\phi} =&\  ((\bm{\lambda}_1\otimes\bm{1}_2^{\mathrm{T}}\otimes\mathcal{T}(\bm{1}_3,3)+\bm{1}_1\otimes\bm{\lambda}_2^{\mathrm{T}}\otimes\ \mathcal{T}(\bm{1}_3,3)+\bm{1}_1\otimes\bm{1}_2^{\mathrm{T}}\otimes\mathcal{T}(\bm{\lambda}_3,3)).^{-1}\\
		&(4\pi\bm{t}\times_1\bm{V}_1^{\mathrm{T}}\times_2\bm{V}_2^{\mathrm{T}}\times_3\bm{V}_3^{\mathrm{T}}))\times_1\bm{V}_1\times_2\bm{V}_2\times_3\bm{V}_3.
	\end{aligned}	
\end{equation}
\begin{tabularx}{\textwidth}{@{}l@{\quad}r@{---}X@{}}
	where & $ ().^{-1}$ & compute reciprocal pointwisely,\\
	&$ \bm{\lambda}_i $ & a column vector composed of the diagonal elements of $ \bm{\Lambda}_i $,\\
	&$ \mathcal{T}(\bm{\lambda}_3,3) $& rotate column vector $\bm{\lambda}_3  $ to the third dimension.
\end{tabularx}
Now, the new solver of the Poisson equation has been completed.

From Eqs.(\ref{Fu}) to (\ref{bijk}), we derive
\begin{equation}\label{gFu}
	\nabla F(\bm{u}) = \sqrt{\dfrac{N_{\mathrm{e}}}{\langle \bm{u},\bm{M}\bullet\bm{u}\rangle}}\left(\nabla E(\bm{u}_{\mathrm{n}},\bm{R})-\dfrac{\bm{u}_{\mathrm{n}}^\mathrm{T} \nabla E(\bm{u}_{\mathrm{n}},\bm{R})}{N_{\mathrm{e}}} \bm{M}\bullet\bm{u}_{\mathrm{n}}  \right)
\end{equation}
where
\begin{equation}\label{gEun}
	\nabla E(\bm{u}_{\mathrm{n}},\bm{R}) = \lambda\bm{A}\bullet\bm{u}_{\mathrm{n}} + \bm{M}\bullet f^{\prime}(\bm{u}_{\mathrm{n}}) + 2\bm{M}\bullet(\bm{u}_{\mathrm{n}}.\cdot\bm{\phi}),
\end{equation}
\begin{tabularx}{\textwidth}{@{}l@{\quad}r@{---}X@{}}
	where & $ f^{\prime}(\bm{u}_{\mathrm{n}})$ & $ f^{\prime}(\bm{u}_{\mathrm{n}})(i,j,k) =  C_{\mathrm{F}}\cdot \frac{10}{3} \bm{u}^{\frac{7}{3}}_{\mathrm{n}}(i,j,k) +\varepsilon_{\mathrm{xc}}^{\prime}(\bm{u}^2_{\mathrm{n}}(i,j,k)) \cdot 2\bm{u}^3_{\mathrm{n}}(i,j,k)+\varepsilon_{\mathrm{xc}}(\bm{u}^2_{\mathrm{n}}(i,j,k))\cdot 2\bm{u}_{\mathrm{n}}(i,j,k)$,\\
	& $ \bm{u}_{\mathrm{n}}.\cdot\bm{\phi}$ & the Hadamard product of $ \bm{u}_{\mathrm{n}}$ and $\bm{\phi} $.
\end{tabularx}

After obtaining $ F(\bm{u}) $ and $ \nabla F(\bm{u}) $, now we propose the nonlinear conjugate gradient algorithm $ \mathsf{u-SG} $ for solving $ \bm{u} $ on a single grid, as shown in Algorithm \ref{uSG} (*denotes user-defined values).
\begin{algorithm}
	\caption{The nonlinear conjugate gradient algorithm $ \mathsf{u-SG} $ for solving $ \bm{u} $ on a single grid.}\label{uSG}
	\begin{algorithmic}[1]
		\Require The parameter sets of the kinetic functional, exchange-correlation functional and the pseudopotential $ \mathscr{P} $, ion system configuration $ \mathscr{C} $, the computational domain $ \Omega $, the mesh size $ h $, the initial guess of $ \bm{u} $ (denoted as $ \bm{u}_0$).
		\Ensure $[\bm{u},f,\bm{g},\bm{\phi}, flag]$ \Comment{$ flag $ is the marker of convergence, 0 denotes non-convergence, 1 denotes convergence.}
		
		\State Compute $ \bm{A} $ and $ \bm{M} $ by $ \Omega $ and $ h $;
		\State Substitute $ \mathscr{P} $ and $ \mathscr{C} $ into Eqs.(\ref{GNH}), (\ref{br}), (\ref{b}) and (\ref{bijk}) to compute the nuclear charge distribution vector $ \bm{b} $;\label{vb}
		\State $ [f_0, \bm{g}_0,\bm{\phi}_0] = \mathsf{FgF}(\bm{u}_0,\bm{b},\mathscr{P},h) $;\Comment{Invoke Algorithm \ref{FgF}}
		\State $ \bm{d}_0 = -\bm{g}_0$;\Comment{The initial search direction}
		\State $ i_{\mathrm{max}}=100 $;  \Comment{*Maximum number of iteration}\label{imax}
		\State $ \varepsilon=1\times 10^{-4} $;\Comment{*The tolerance of the relative error}
		\State $ flag==0 $;
		\For{$k=0:i_{\mathrm{max}}-1$}
		\State $[\bm{u}_{k+1},f_{k+1},\bm{g}_{k+1},\bm{\phi}_{k+1}] = \mathsf{LineSearch}(\bm{u}_{k}, \bm{d}_k,  f_{k},\bm{g}_{k},\bm{b},\bm{A},\bm{M},\mathscr{P},h)$;\Comment{Line search, invoke Algorithm \ref{LineSearch}}\label{LS}
		\State$ \varepsilon_{\mathrm{n}}=\dfrac{| f_{k+1} - f_{k}|}{N_{\mathrm{a}}} $;\Comment{The relative error, where $ N_{\mathrm{a}} $ denotes the number of atoms}
		\If{$ \varepsilon_{\mathrm{n}}<\varepsilon $}
		\State $ \bm{u}=\bm{u}_{k+1} $;
		\State $ f=f_{k+1} $;
		\State $ \bm{g}=\bm{g}_{k+1} $;
		\State $ \bm{\phi}=\bm{\phi}_{k+1} $;
		\State $ flag=1 $;
		\State break
		\EndIf
		\State $ \beta_{k+1} =\dfrac{\langle \bm{g}_{k+1}, \bm{g}_{k+1} \rangle - \langle \bm{g}_{k+1}, \bm{g}_{k} \rangle}{\langle \bm{g}_{k}, \bm{g}_{k} \rangle} $; \Comment{Polak-Ribiere-Polyak rule}		
		
		\algstore{bkbreak}
	\end{algorithmic}
\end{algorithm}
\begin{algorithm}
\begin{algorithmic}
\algrestore{bkbreak}				
		
		\State $ \bm{d}_{k+1}=-\bm{g}_{k+1} + \beta_{k} \bm{d}_{k} $;\Comment{The updated search direction}
		\State $ \bm{u}_{k}=\bm{u}_{k+1} $;
		\State $ f_{k}=f_{k+1} $;
		\State $ \bm{g}_{k}=\bm{g}_{k+1} $;
		\State $ \bm{d}_{k}=\bm{d}_{k+1} $;    
		\State $ \bm{\phi}_{k}=\bm{\phi}_{k+1} $.		
		\EndFor
	\end{algorithmic}
\end{algorithm}
$ \mathsf{LineSearch} $ in the \ref{LS}th row of the algorithm is a line search function for solving the optimization problem in Eq.(\ref{alpha}), and simultaneously provide the values of $ \bm{u}_{k+1} $, $ f_{k+1} $, $ \bm{g}_{k+1}$ and $ \bm{\phi}_{k+1} $ when $ \alpha=\alpha^* $.
\begin{equation}\label{alpha}
	\alpha^* = \arg\min_{\alpha\geqslant 0}(F(\bm{u}_{k}+\alpha\bm{d}_{k})).
\end{equation}
Methods in current literature all used exact line search, and did not provide detailed steps. Exact line search is not only computationally expensive, but also unnecessary, since the algorithm is to solve the minimum value of $ F(\bm{u}) $ in the admissible set,  we only need to compute an iteration point where $ F(\bm{u}) $ has a sufficiently large value of decreasing at current search direction. \emph{By using the Wolfe rule (an inexact line search rule) and an improved bisection method, this work proposes an inexact line search function $ \mathsf{LineSearch} $ of which detailed steps are shown in Algorithm \ref{LineSearch}.}
\begin{algorithm}
	\caption{The linear search function $ \mathsf{LineSearch} $}\label{LineSearch}	
	\begin{algorithmic}[1]
		\Require $ \bm{u}_k, \bm{d}_k, f_k, \bm{g}_k, \bm{b},\bm{A},\bm{M}, \mathscr{P}, h$
		\Ensure $[\bm{u}_{k+1}, f_{k+1}, \bm{g}_{k+1}, \bm{\phi}_{k+1}] $
		
		\Comment{Explore the interval $ [\alpha_-,\alpha_{+}] $ in which $ \alpha_* $ lies}
		\State $ p_k = \langle \bm{g}_k, \bm{d}_k \rangle$;
		\State $ \alpha_-=0 $;
		\State $ \alpha_{+}=1 $;\Comment{*The initial step length}
		\State $ p_- =p_k$;
		\If{$ p_->0 $}\Comment{$ \alpha^* =0$}
		\State $ \bm{u}_{k+1}=\bm{u}_k $;
		\State $ f_{k+1}=f_k $;
		\State $ \bm{g}_{k+1}=\bm{g}_k $;
		\State $ \bm{\phi}_{k}=\bm{\phi}_{k+1} $;
		\State return
		\EndIf
		\State $ [p_+, \bm{u}_{k+1}, f_{k+1}, \bm{g}_{k+1}, \bm{\phi}_{k+1}] = \mathsf{der}(\bm{u}_k,  \alpha_+, \bm{d}_k, \bm{b}, \bm{A},\bm{M},\mathscr{P}, h)$;\Comment{Invoke Algorithm \ref{der}}
		\If{$\mathsf{WolfeCheck}(\alpha_+, \bm{d}_k, f_{k+1}, \bm{g}_{k+1} ,f_{k}, p_{k})$}
		\State return
		\EndIf
		\State $ f_{\mathrm{o}}=f_{k+1} $;
		\State $ j_{\mathrm{max}}=30 $;\Comment{*Maximum number of exploration}
		\State $ flag=0 $;\Comment{The marker of exploration result}
		\For{$ j=1:j_{\mathrm{max}} $}
		
		\algstore{bkbreak}
	\end{algorithmic}
\end{algorithm}
\begin{algorithm}
\begin{algorithmic}
\algrestore{bkbreak}				
		
		\If{$ p_+>0 $}
		\State $ flag=1 $;\Comment{Exploration succeeded.}
		\State break
		\Else
		\State $ \alpha_-=\alpha_{+} $;\Comment{Regarding the end point of the previous interval as the new starting point.}
		\State $ \alpha_+=2\alpha_{+} $;\Comment{Double the exploration step.}
		\State $ p_- =p_+ $;
		\State $ [p_+, \bm{u}_{k+1}, f_{k+1}, \bm{g}_{k+1}, \bm{\phi}_{k+1}] = \mathsf{der}(\bm{u}_k,  \alpha_+, \bm{d}_k, \bm{b}, \bm{A},\bm{M},\mathscr{P}, h)$;   
		\If{$\mathsf{WolfeCheck}(\alpha_+, \bm{d}_k, f_{k+1}, \bm{g}_{k+1} ,f_{k}, p_{k})\ \mathrm{or}\  f_{\mathrm{o}}<f_{k+1}$}
		\State return		
		\EndIf
		\State $ f_{\mathrm{o}} =f_{k+1}$;		
		\EndIf  
		\EndFor		
		\Comment{Searching $ \alpha_* $ in interval $ [\alpha_-,\alpha_{+}] $}
		\State $ l=\alpha_+-\alpha_- $;
		\State $ \varepsilon=1\times 10^{-4} $;\Comment{*The tolerance of the relative error.}
		\If{$ flag ==1$}
		\State $ i_{\mathrm{max}}=30 $;\Comment{*The maximum number of searching in the interval.}
		\For{$ j=1:i_{\mathrm{max}} $}
		\If{$ \frac{\alpha_+-\alpha_-}{l} < \varepsilon$}
		\State break
		\EndIf
		\State $ x= \frac{\alpha_+-\alpha_-}{2}$;
		\State $ [p_x, \bm{u}_{k+1}, f_{k+1}, \bm{g}_{k+1}, \bm{\phi}_{k+1}] = \mathsf{der}(\bm{u}_k,  x, \bm{d}_k, \bm{b},\bm{A},\bm{M}, \mathscr{P}, h)$;
		\If{$\mathsf{WolfeCheck}(x, \bm{d}_k, f_{k+1}, \bm{g}_{k+1} ,f_{k}, p_{k})$}
		\State return
		\EndIf
		\If{$ p_x>0 $}
		\State $ x_0=\alpha_--\frac{(\alpha_--x)p_-}{p_--p_x} $;
		\State $ [p_{x_0}, \bm{u}_{k+1}, f_{k+1}, \bm{g}_{k+1}, \bm{\phi}_{k+1}] = \mathsf{der}(\bm{u}_k,  x_0, \bm{d}_k, \bm{b}, \bm{A},\bm{M},\mathscr{P}, h)$;
		\If{$\mathsf{WolfeCheck}(x_0, \bm{d}_k, f_{k+1}, \bm{g}_{k+1} ,f_{k}, p_{k})$}
		\State return
		\EndIf
		\If{$ p_{x_0}>0 $}
		\State $\alpha_{+}=x_0$;
		\State $ p_+=p_{x_0} $;
		\Else
		\State $ \alpha_-=x_0 $;
		\State $ \alpha_{+}=x $;
		\State $ p_- =p_{x_0}$;
		\State $ p_+=p_x $;
		\EndIf	
		\Else  		   
		\State $ x_0=\alpha_+-\frac{(\alpha_+-x)p_{+}}{p_{+}-p_x} $;
		\State $ [p_{x_0}, \bm{u}_{k+1}, f_{k+1}, \bm{g}_{k+1}, \bm{\phi}_{k+1}] = \mathsf{der}(\bm{u}_k,  x_0, \bm{d}_k, \bm{b}, \bm{A},\bm{M},\mathscr{P}, h)$;
		
				\algstore{bkbreak}
	\end{algorithmic}
\end{algorithm}
\begin{algorithm}
\begin{algorithmic}
\algrestore{bkbreak}
		
		\If{$\mathsf{WolfeCheck}(x_0, \bm{d}_k, f_{k+1}, \bm{g}_{k+1} ,f_{k}, p_{k})$}
		\State return
		\EndIf	
		\If{$ p_{x_0}>0 $}
		\State $ \alpha_{-}=x $;
		\State $ \alpha_{+}=x_0 $;
		\State $ p_- =p_{x}$;
		\State $ p_+=p_{x_0} $;
		\Else
		\State $\alpha_{-}=x_0$;
		\State $ p_-=p_{x_0} $;  
		\EndIf      
		\EndIf   
		\EndFor
		\State $ \alpha_*=\frac{\alpha_{-}+\alpha_{+}}{2} $;
		\State $ [p_{\alpha}, \bm{u}_{k+1}, f_{k+1}, \bm{g}_{k+1}, \bm{\phi}_{k+1}] = \mathsf{der}(\bm{u}_k,  \alpha, \bm{d}_k, \bm{b}, \bm{A},\bm{M},\mathscr{P}, h)$;
		\EndIf   
	\end{algorithmic}
\end{algorithm}
where function $ \mathsf{der} $ (see Algorithm \ref{der}) is for computing the derivative $ F(\bm{u}_{k}+\alpha\bm{d}_{k}) $at $ \alpha $, and simultaneously provide the values of $ \bm{u}_{k+1}$, $f_{k+1}$, $\bm{g}_{k+1}$ and $ \bm{\phi}_{k+1} $. Function $ \mathsf{WolfeCheck} $ (see Algorithm \ref{WolfeCheck}) is for checking whether $ \alpha $ satisfies the Wolfe rule. Each iteration of $ \bm{u} $ in Algorithm \ref{der} need to compute $ F(\bm{u}) $ and $ \nabla F(\bm{u}) $ once, thus, in order to reduce the number of times of computing the Poisson equation from two to one, the computation of the two values can be integrated into one function, as shown in Algorithm \ref{FgF}.
\begin{algorithm}
	\caption{Function $ \mathsf{der} $}\label{der}
	\begin{algorithmic}[1]
		\Require $ \bm{u}_k,  \alpha, \bm{d}_k, \bm{b}, \bm{A},\bm{M},\mathscr{P}, h$
		\Ensure  $[p_{k+1}, \bm{u}_{k+1}, f_{k+1}, \bm{g}_{k+1}, \bm{\phi}_{k+1}] $
		
		\State $\bm{u}_{k+1}=\bm{u}_{k} + \alpha\bm{d}_{k}  $\;
		\State $ [f_{k+1}, \bm{g}_{k+1},\bm{\phi}_{k+1}] = \mathsf{FgF}(\bm{u}_k,\bm{b},\bm{A},\bm{M},\mathscr{P},h) $\Comment{Invoke Algorithm \ref{FgF}}
		\State $ p_{k+1} =  \langle \bm{g}_{k+1}, \bm{d}_k \rangle $.\Comment{$ \frac{\mathrm{d}F}{\mathrm{d}\alpha}$}
	\end{algorithmic}	
\end{algorithm}
\begin{algorithm}
	\caption{Function $ \mathsf{WolfeCheck} $}\label{WolfeCheck}
	\begin{algorithmic}[1]
		\Require $ \alpha, \bm{d}_k, f_{k+1}, \bm{g}_{k+1} ,f_{k}, p_{k} $
		\Ensure Logical value $a $
		
		\State $ c_1=0.01 $;\Comment{*A parameter}
		\State $ c_2=0.1 $;\Comment{*A parameter, with $ 0<c_1<c_2<\frac{1}{2} $}
		\State $ p_{k+1} =  \langle \bm{g}_{k+1}, \bm{d}_k \rangle $;
		\State $ a=(f_{k+1} \leqslant f_{k} + c_1 \alpha p_k )\ \mathrm{and}\ (| p_{k+1} | \leqslant c_2 | p_k |) $.
	\end{algorithmic}
\end{algorithm}
\begin{algorithm}
	\caption{Function $ \mathsf{FgF} $}\label{FgF}
	\begin{algorithmic}[1]
		\Require $ \bm{u}, \bm{b},\bm{A},\bm{M},  \mathscr{P}, h$ 
		\Ensure $[f, \bm{g}, \bm{\phi}] $
		
		\State	$ \bm{u}_{\mathrm{n}} = \bm{u}\sqrt{\dfrac{N_{\mathrm{e}}}{\langle \bm{u},\bm{M}\bullet\bm{u}\rangle}}$;
		\State $ \bm{t}=\bm{M}\bullet\bm{u}_{\mathrm{n}} + \bm{b} $;
		\State Substitute $ \bm{t} $ into Eq.(\ref{phisol}) to compute $ \bm{\phi} $;
		\State Substitute $ \bm{A} $, $ \bm{\phi} $ and $\mathscr{P}  $ into Eq.(\ref{Eun}) to compute $f= F(\bm{u}) $;
		\State Substitute $ \bm{A} $, $ \bm{\phi} $ and $\mathscr{P}  $ into Eq.(\ref{gEun}) to compute $ \bm{g}=\nabla F(\bm{u}) $.
	\end{algorithmic}
\end{algorithm}

Lastly, we need to compute $ E_{\mathrm{c}}(\bm{R}) $. According to Eq.(\ref{phic}), let $ \bm{t}=\tilde{\bm{b}}- \bm{b} $
, then from Eq.(\ref{phisol}) we derive $ \bm{\phi}_{\mathrm{c}} $, and by substituting into Eq.(\ref{Ec2}), we derive
\begin{equation}\label{Ec2d}
	E_{\mathrm{c}}(\bm{R})=\dfrac{1}{2}\left(\langle \tilde{\bm{b}}+\bm{b}, \bm{\phi}_{\mathrm{c}}\rangle+ \sum_{J=1}^{M} \langle \bm{b}_{J}, \bm{\phi}_{J}\rangle-\sum_{J=1}^{M} \langle \tilde{\bm{b}}_{J}, \tilde{\bm{\phi}}_{J}\rangle\right).
\end{equation}

Finally, the total energy is derived as 
\begin{equation}\label{Eunf}
	E(\bm{u}_{\mathrm{n}},\bm{R}) = \dfrac{\lambda}{2}\langle \bm{u}_{\mathrm{n}}, \bm{A}\bullet\bm{u}_{\mathrm{n}}\rangle + \left(\sum f(\bm{u}_{\mathrm{n}})\right)\cdot h^3 + \dfrac{1}{2}(\langle \bm{M}\bullet\bm{u}_{\mathrm{n}}.^2+ \bm{b}, \bm{\phi}\rangle) - \dfrac{1}{2} \sum_{J=1}^{M} \langle \bm{b}_{J}, \bm{\phi}_{J}\rangle+E_{\mathrm{c}}(\bm{R}).
\end{equation}

\section{The multi-grid solver}
The number of iterations needed for convergence in the single-grid algorithm rapidly increases with the number of elements. Meanwhile, for the nonlinear conjugate gradient method, finding a proper preconditioner is very difficult. To improve the computational efficiencies, following the idea of \cite{Nash2000} and the proposed single-grid algorithm, this section proposes a multi-grid algorithm $ \mathsf{u-MG} $ for solving $ \bm{u} $ (the number of levels is set as three in this work), as shown in Algorithm \ref{algMG}.
\begin{algorithm}
	\caption{The multi-grid algorithm $\mathsf{ u-MG}$ for solving $ \bm{u} $.}\label{algMG}
	\begin{algorithmic}[1]
		\Require $ \mathscr{P}, \mathscr{C}, \Omega $, the mesh size of level 0 $ h_{[0]} $, the initial guess of $ \bm{u} $ at level 2 (denotes $ \bm{u}_{[2],0} $), the vector of ion charge distribution at level 0 $\bm{b}_{[0]}$
		\Ensure $[\bm{u}_{[0]},f_{[0]}, \bm{g}_{[0]}, \bm{\phi}_{[0]}] $
		
		\Comment{For a single-atom system, computing the initial guess of $ \bm{u} $ at level 0 by interpolation level-by-level; for a multi-atom system, the following steps can be replaced by the superposition of single-atom electron density.}
		\State $ h_{[2]}=4h_{[0]} $;
		\State $ h_{[1]}=2h_{[0]} $;
		\State $ [\bm{u}_{[2]},f_{[2]},\bm{g}_{[2]},\bm{\phi}_{[2]},\ flag]=\mathsf{u-SG}(\mathscr{P},\mathscr{C},\Omega,h_{[2]}, \bm{u}_{[2],0}) $;\Comment{Invoke Algorithm \ref{uSG}}
		\State $ \bm{u}_{[1],0}= I_{h_{[2]}}^{h_{[1]}}\bm{u}_{[2]}$;
		\State $ [\bm{u}_{[1]},f_{[1]},\bm{g}_{[1]},\bm{\phi}_{[1]},\ flag]=\mathsf{u-SG}(\mathscr{P},\mathscr{C},\Omega,h_{[1]}, \bm{u}_{[1],0}) $;
		
		\algstore{bkbreak}
	\end{algorithmic}
\end{algorithm}
\begin{algorithm}
\begin{algorithmic}
\algrestore{bkbreak}		
		
		\State $ \bm{u}_{[0],0}= I_{h_{[1]}}^{h_{[0]}}\bm{u}_{[1]}$;	
		
		\Comment{Begin multi-grid iteration}
		\State $ \bm{b}_{[1]}= I_{h_{[0]}}^{h_{[1]}}\bm{b}_{[0]} $;
		\State $ \bm{b}_{[2]}= I_{h_{[1]}}^{h_{[2]}}\bm{b}_{[1]} $;
		\State Compute $ \bm{A}_{[0]}, \bm{M}_{[0]}, \bm{A}_{[1]}, \bm{M}_{[1]}$;\Comment{Preparing for the line search at rows \ref{LS1} and \ref{LS0}.}
		\State $ i_{\mathrm{max}}=20 $;\Comment{*The maximum iteration number.}
		\State $ j_{\mathrm{max}}=[5\ 5\ 5] $;\Comment{*The maximum iteration number of each level.}
		\State $ flag=0 $;\Comment{The marker of convergence of level 0, 0 denotes convergence failure, 1 denotes convergence.}
		\For {$ i=1: i_{\mathrm{max}}$}
		\For {$ l=0:2 $}\Comment{For each level}
		\Switch{$ l $}
		\Case{0}		
		\State Let $ \tilde{F}_{[l]}(\bm{u}) = F_{[l]}(\bm{u})$;\Comment{$ \tilde{F}_{[l]}(\bm{u}) $denotes the actual objective function at level $ l $, $ F_{[l]}(\bm{u}) $ denotes $ F(\bm{u}) $ at level $ l $}
		\State Substitute $ F(\bm{u}) $ in Algorithm \ref{FgF} with $\tilde{F}_{[l]}(\bm{u})  $; substitute row \ref{vb} in Algorithm \ref{uSG} with $ \bm{b}_{[l]} $, assign $ j_{\mathrm{max}}(l) $ at row \ref{imax}, take $ \bm{u}_{[l],0} $ as the initial guess, proceed Algorithm \ref{uSG} and obtain $ [\bm{u}_{[l]},f_{[l]},\bm{g}_{[l]},\bm{\phi}_{[l]},flag] $;
		\If{$ flag==1 $}
		\State break
		\EndIf
		\EndCase
		\Case{1}	
		\State $ \bm{u}_{[1],0}= I_{h_{[0]}}^{h_{[1]}}\bm{u}_{[0]} $;
		\State $ \bm{g}_{[1]}=\nabla F_{[1]}(\bm{u}_{[1],0}) $;
		\State $ \bm{v}_1 = \bm{g}_{[1]}- I_{h_{[0]}}^{h_{[1]}}\bm{g}_{[0]}$;
		\State Let $ \tilde{F}_{[l]}(\bm{u}) = F_{[l]}(\bm{u})-\langle \bm{v}_1, \bm{u}\rangle$;
		\State Substitute $ F(\bm{u}) $ in Algorithm \ref{FgF} with $\tilde{F}_{[l]}(\bm{u})  $; substitute row \ref{vb} in Algorithm \ref{uSG} with $ \bm{b}_{[l]} $, take $ \bm{u}_{[l],0} $ as the initial guess, proceed Algorithm \ref{uSG}, skip the judgment of the iteration error, iterate $ j_{\mathrm{max}}(l) $ time and obtain $ [\bm{u}_{[l]},f_{[l]},\bm{g}_{[l]},\bm{\phi}_{[l]},flag] $;
		\EndCase		
		\Case{2}		
		\State $ \bm{u}_{[2],0}= I_{h_{[1]}}^{h_{[2]}}\bm{u}_{[1]} $;
		\State $ \bm{g}_{[2]}=\nabla F_{[2]}(\bm{u}_{[2],0})-\langle 8I_{h_{[1]}}^{h_{[2]}}\bm{v}_1, \bm{u}_{[2],0}\rangle $;\Comment{Multiplying $ I_{h_{[1]}}^{h_{[2]}} $ with $ 8=2^3 $ is to preserve the inner product.}				
		\State $ \bm{v}_2 = \bm{g}_{[2]}- I_{h_{[1]}}^{h_{[2]}}\bm{g}_{[1]}$;
		\State Let $ \tilde{F}_{[l]}(\bm{u}) = F_{[l]}(\bm{u})-\langle 8I_{h_{[1]}}^{h_{[2]}}\bm{v}_1, \bm{u}\rangle - \langle \bm{v}_2, \bm{u}\rangle$;
		\State Substitute $ F(\bm{u}) $ in Algorithm \ref{FgF} with $\tilde{F}_{[l]}(\bm{u})  $; substitute row \ref{vb} in Algorithm \ref{uSG} with $ \bm{b}_{[l]} $, take $ \bm{u}_{[l],0} $ as the initial guess, proceed Algorithm \ref{uSG}, skip the judgment of iteration number, iterate $ j_{\mathrm{max}}(l) $ times and obtain $ [\bm{u}_{[l]},f_{[l]},\bm{g}_{[l]},\bm{\phi}_{[l]},flag] $;
		\EndCase
		\EndSwitch			
		\EndFor	
		
		\algstore{bkbreak}
	\end{algorithmic}
\end{algorithm}
\begin{algorithm}
\begin{algorithmic}
\algrestore{bkbreak}
		
		\If{$ flag==1 $}
		\State break\Comment{Computation completed.}
		\EndIf				
		\State $ \bm{e}_{[1]}= \bm{I}_{h_{[2]}}^{h_{[1]}}(\bm{u}_{[2]}-\bm{u}_{[2],0}) $;
		\State Substitute $ F(\bm{u}) $ in Algorithm \ref{FgF} with $\tilde{F}_{[1]}(\bm{u})  $, take $ \bm{u}_{[1]} $ as the starting point and $ \bm{e}_{[1]} $ as the search direction, proceed Algorithm \ref{LineSearch} and obtain $ [\bm{u}_{[1]},f_{[1]},\bm{g}_{[1]},\bm{\phi}_{[1]}] $;\label{LS1}
		\State $ \bm{e}_{[0]}= \bm{I}_{h_{[1]}}^{h_{[0]}}(\bm{u}_{[1]}-\bm{u}_{[1],0}) $;
		\State Substitute $ F(\bm{u}) $ in Algorithm \ref{FgF} with $\tilde{F}_{[0]}(\bm{u}) $, take $ \bm{u}_{[0]} $ as the starting point and $ \bm{e}_{[0]} $ as the searching direction, proceed Algorithm \ref{LineSearch} and obtain $ [\bm{u}_{[0]},f_{[0]},\bm{g}_{[0]},\bm{\phi}_{[0]}] $;\label{LS0}
		\State $ \bm{u}_{[0],0}=\bm{u}_{[0]} $.
		\EndFor		
	\end{algorithmic}
\end{algorithm}
where $ I_H^h $ denotes the prolongation operator from the coarse grid with size $ H $ to the fine grid with size $ h $, $ I_h^H $ denotes the restriction operator from the fine grid with size$ h $ to the coarse grid with size $ H $, as shown in Eq. (\ref{I12}),
\begin{subequations}\label{I12}
	\begin{alignat}{1}
		I_H^h \bm{v}_H=\ &\ \bm{v}_H \times_1 \bm{L}_1 \times_2 \bm{L}_2 \times_3 \bm{L}_3,\\
		I_h^H \bm{v}_h=\ &\ \bm{v}_h \times_1 \frac{1}{2}\bm{L}_1^{\mathrm{T}} \times_2 \frac{1}{2}\bm{L}_2^{\mathrm{T}} \times_3 \frac{1}{2}\bm{L}_3^{\mathrm{T}},
	\end{alignat}
\end{subequations}
where the uni-dimensional prolongation matrix
\begin{equation}
	\bm{L}_i=
	\dfrac{1}{2}\begin{bmatrix}
		1 & 2 & 1 & & & & \\
		& & 1 & 2 & 1 & & \\
		& & &\cdots\\
		& & & & 1 & 2 & 1
	\end{bmatrix}^{\mathrm{T}},\quad i=1,2,3
\end{equation}
has proper orders.

\section{Numerical examples}
For the sake of convenience, all numerical examples in this work only use atom clusters of aluminum. According to \cite{GarciaCervera2007}, the parameters of the pseudopotential in Eq.(\ref{GNH}) are set as follows: $Z=3 $, $ r_{\mathrm{c}}=3.5 $, $ A=0.1107$, $ R=1.150 \ $Bohr.
\subsection{Single-atom systems}
\begin{figure}[htpb]
	\centering
	\includegraphics[width = 0.6\textwidth]{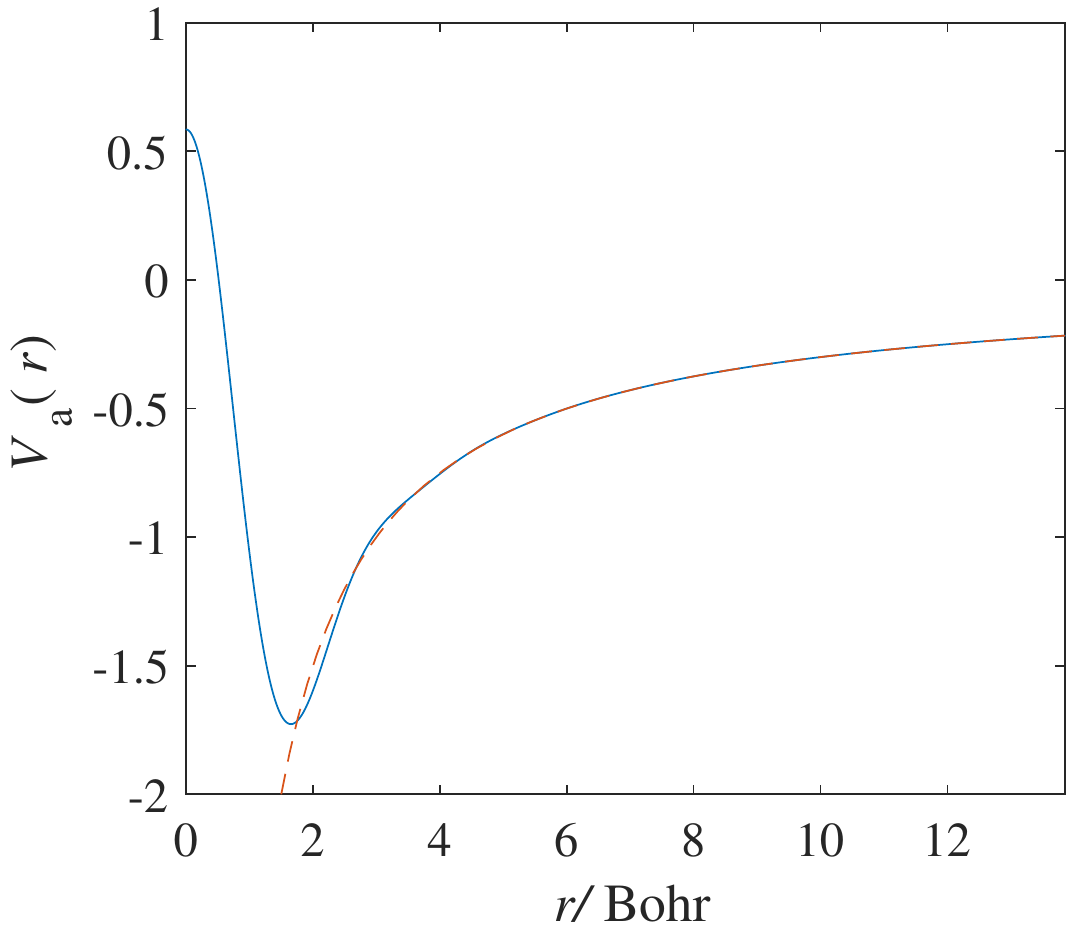}
	\caption{The figure of $ V_{\mathrm{a}}(r) $}\label{Vrfig}
\end{figure}

Firstly, single-atom systems are considered. According to Algorithm \ref{uSG}, to compute the electron density of the system, four parameters need to be selected: the order of finite difference $ n $, radius $ R_\mathrm{a} $, the mesh size at level 0 $ h_{[0]} $ and the tolerance of the relative error $ \varepsilon $. On the one hand, to ensure accuracies, $ R_\mathrm{a} $ should be sufficiently large, and $ h_{[0]} $ and $ \varepsilon $ should be sufficiently small; on the other hand, to balance the efficiencies, $ R_\mathrm{a} $should be sufficiently small, and $ h_{[0]} $ and $ \varepsilon $ should be sufficiently large. Thus, the parameters should be selected according to numerical experiments. The figure of $ V_{\mathrm{a}}(r) $ can be derived from Eq.(\ref{GNH}), as shown in Fig.\ref{Vrfig} where the dashed line represents the figure of $ -3/r $.

\begin{figure}[htpb]
	\centering
	\includegraphics[width = 0.6\textwidth]{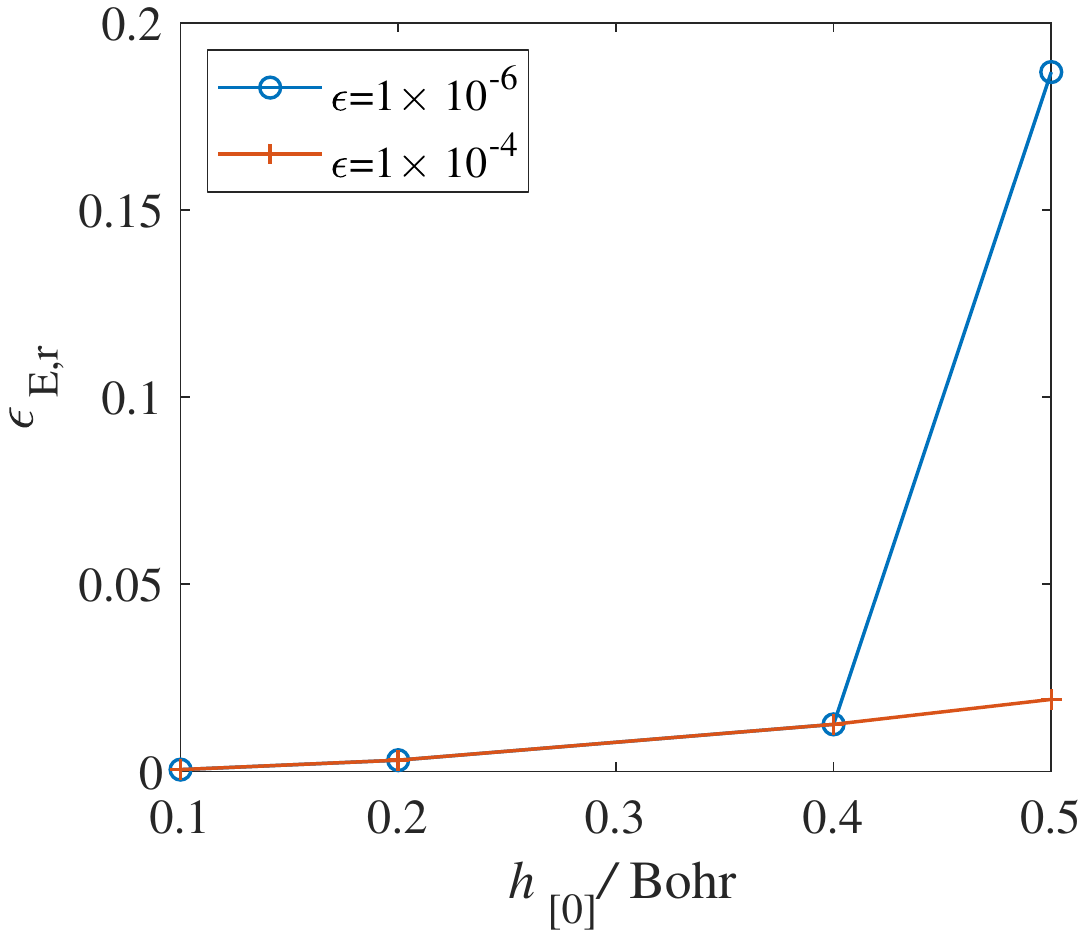}
	\caption{The figure of $ \varepsilon_{\mathrm{E,r}} $ of the first group of experiment of a single atom.}\label{2_ex1}
\end{figure}

Firstly, we set $ \varepsilon = 1\times 10^{-8}$, $ h_{[0]} = 0.0667 $ Bohr, and computed the reference solution of the total energy $E_{\mathrm{ref}}=\ -58.0071 $ eV by using Algorithm \ref{algMG} and Eq.(\ref{Eunf}). Then, we set $ n=1 $, $ R_\mathrm{a}=8 $Bohr, and proceeded the first group of experiment by setting $( \varepsilon, h_{[0]})$ as $ [1\times 10^{-6},1\times 10^{-4}] \times [0.1, 0.2, 0.4, 0.5]$. For each combination of parameters, we computed the relative error
\begin{equation}
	\varepsilon_{\mathrm{E,r}}=\dfrac{E_{\mathrm{ref}}-E}{E_{\mathrm{ref}}},
\end{equation}
the results are illustrated in Fig.\ref{2_ex1}.

Fig.\ref{2_ex1}illustrates that except the case where $( \varepsilon, h_{[0]})=(1\times 10^{-6},\ 0.5)$, the curves of different values of $ \varepsilon $ almost coincide, indicating that in the selected interval, the value of $ \varepsilon $ has little effect on the accuracies of the total energy in most cases, the accuracy with $ \varepsilon = 1\times 10^{-4}$ absolutely reaches the level of accuracy with $ \varepsilon = 1\times 10^{-8}$. For a fixed $ \varepsilon $, $ \varepsilon_{\mathrm{E,r}} $ increases almost linearly with $ h_{[0]} $. When $ ( \varepsilon, h_{[0]})=(1\times 10^{-4},\ 0.5) $, $ \varepsilon_{\mathrm{E,r}} = 0.019$.When $( \varepsilon, h_{[0]})=(1\times 10^{-6},\ 0.5)$, the iteration at level 2 did not converge, leading to the early termination of iteration at level 0.

\begin{figure}[htpb]
	\centering
	\includegraphics[width = 0.6\textwidth]{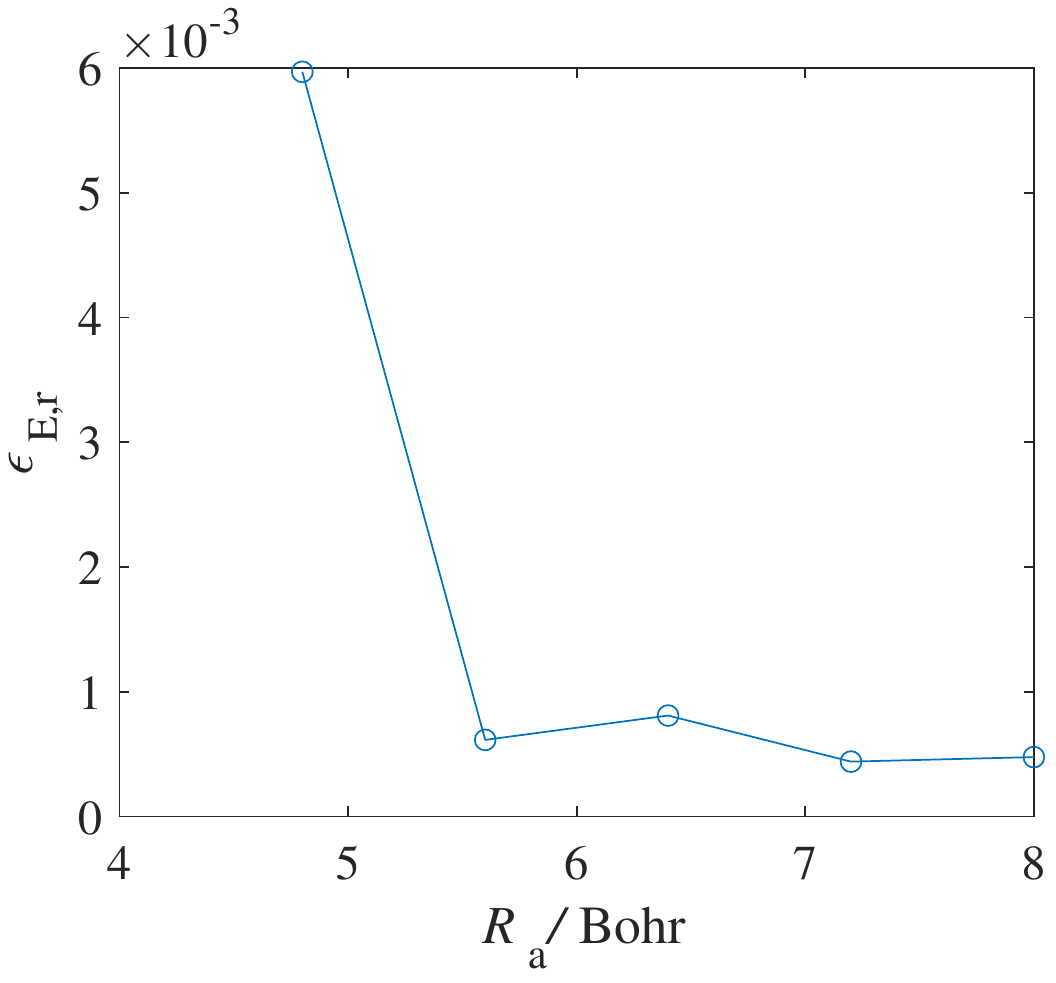}
	\caption{The figure of $ \varepsilon_{\mathrm{E,r}} $ of the second group of experiment of a single atom.}\label{2_ex2}
\end{figure}

Based on the results above, to examine to effect of $ R_{\mathrm{a}} $ on $ \varepsilon_{\mathrm{E,r}} $, we set $ \varepsilon=1\times 10^{-4} $,  and proceeded the second group of experiment. By setting$ (R_{\mathrm{a}},h_{[0]}) $ as $ [8, 7.2, 6.4, 5.6, 4.8]\times 0.1$, we computed the values of $ \varepsilon_{\mathrm{E,r}} $ (the results of $ R_{\mathrm{a}} =8$ are obtained from the first group of experiment), and the results are illustrated in Fig.\ref{2_ex2}.

Fig.\ref{2_ex2} indicates that the values of $ \varepsilon_{\mathrm{E,r}} $ with $ R_{\mathrm{a}} =4.8 $Bohr are apparently higher than that with $ R_{\mathrm{a}}\geqslant 5.6 $, and the values of $ \varepsilon_{\mathrm{E,r}} $ with $ R_{\mathrm{a}}\geqslant 5.6 $ are always smaller than $ 8\times 10^{-4} $, indicating that ensuring accuracies requires $ R_{\mathrm{a}}\geqslant 5.6 $.

\begin{figure}[htpb]
	\centering
	\includegraphics[width = 0.6\textwidth]{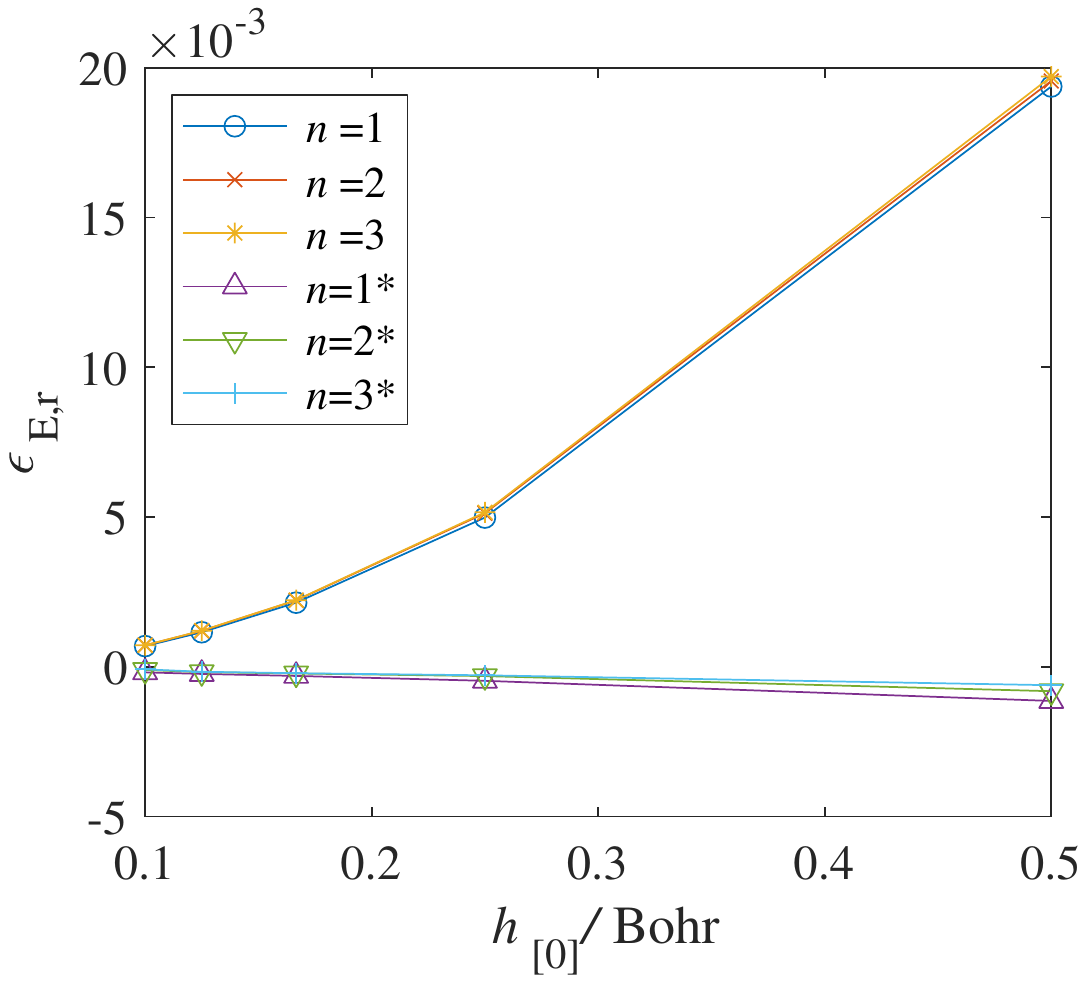}
	\caption{The figure of $ \varepsilon_{\mathrm{E,r}} $ of the third group of experiment of a single atom.}\label{2_ex3}
\end{figure}

Based on the results above, by setting $ \varepsilon=1\times 10^{-4}$, $ R_{\mathrm{a}}=6\ $Bohr, $ h_{[0]}=[0.1,\ 0.1875,\ 0.3,\ 0.5] $Bohr, we proceeded six groups of experiment to examine the effect of different finite difference schemes on computational accuracies. In the first three groups of experiment, we set $ n = 1, 2, 3$ in Eq.(\ref{wp})  and set $ \bm{Q}_i $ in Eq.(\ref{ABQC}) as the integral scheme (named as the Simpson scheme); in the other three groups of experiment,  we set $ n = 1, 2, 3$ in Eq.(\ref{wp}) and set the identity matrix as the integral scheme (named as the zero-order scheme). The results are illustrated in Fig.\ref{2_ex3} (where '*' denotes the second three groups of experiment).

Fig.\ref{2_ex3} illustrates that relative errors with different finite difference schemes all increases with the mesh size. The relative errors with Simpson scheme are always positive and increase rapidly to about 0.02 when $ h_{[0]}=0.5 $Bohr, while the relative errors with the zero-order scheme are always negative and increase only to about 0.001 when $ h_{[0]}=0.5 $Bohr. In addition, for each integral scheme, the curves of the relative errors with different finite difference schemes almost coincide, indicating that for a fixed integral scheme, different finite difference orders have little impact on the accuracies, nonetheless, the group with $ n=3^* $ is slightly more accurate than other groups.

In summary, we suggest selecting $ n=3$, the zero-order integral scheme, $ \varepsilon = 1\times 10^{-4} $, $ R_{\mathrm{a}}=6 $ Bohr. The proper value of $ h_{[0]} $ should be determined according to the results of multi-atom systems. 

\subsection{Multi-atom systems}
For multi-atom systems, $ 1\times 1\times 1 $ face-centered cubic cell is considered first. The initial guess of the covalent electron density is assumed to be the superposition of the covalent electron density of a single atom. Let the lattice constant $ a_0=8\ $Bohr, the reference value of the atom-averaged ground state energy is -59.2280 eV\cite{Suryanarayana2014}.

\begin{figure}[htpb]
	\centering
	\includegraphics[width = 0.6\textwidth]{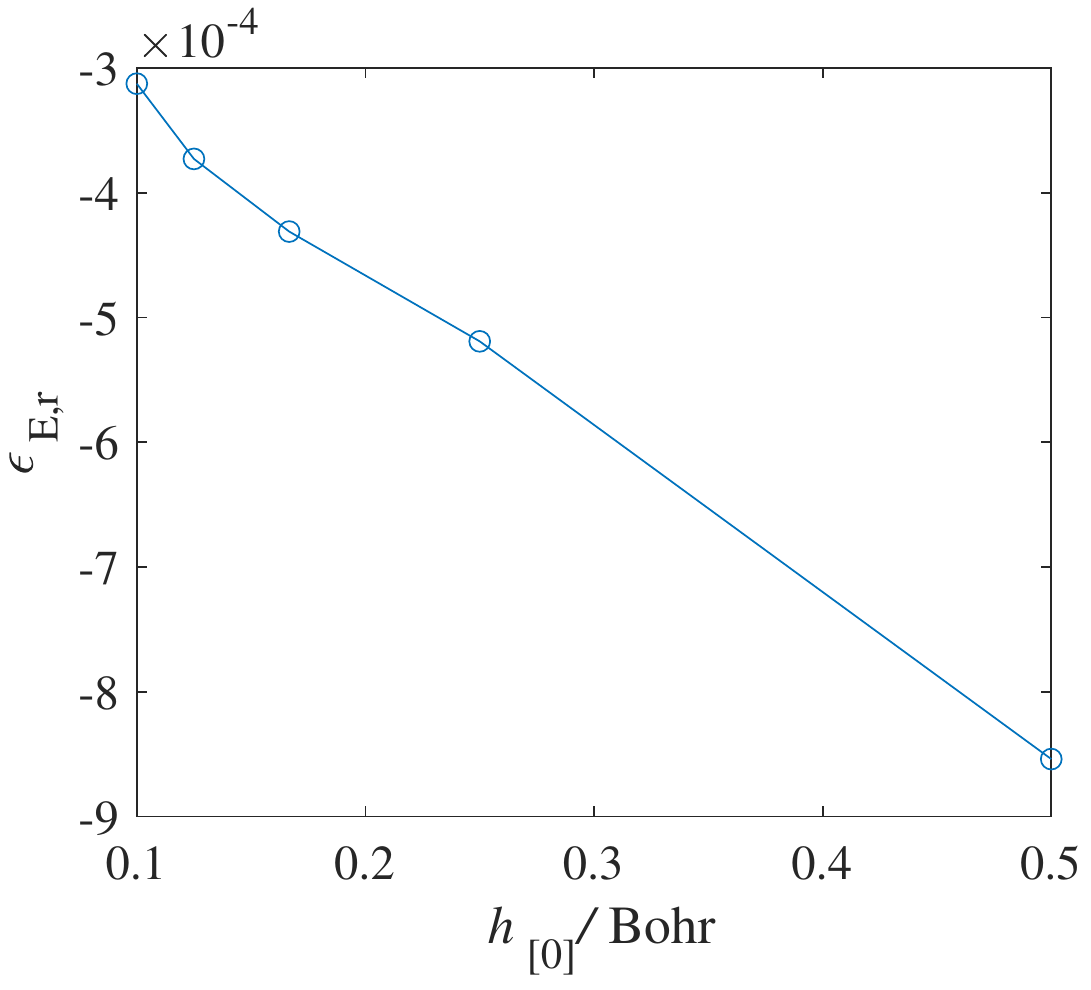}
	\caption{The figure of relative error-mesh size relationship of the $ 1\times 1\times 1 $ unit cell.}\label{2_ex5}
\end{figure}

By setting $ h_{[0]}=[1/10,\ 1/8,\ 1/6,\ 1/4,\ 1/2] $Bohr, we proceeded the first group of experiment and obtained the results shown in Fig.\ref{2_ex5}.
Fig.\ref{2_ex5} indicates that the values of $ \varepsilon_{\mathrm{E,r}} $ are always negative and decreases with $ h_{[0]} $, the minimum value is about $ 8.5\times 10^{-4} $.

To investigate the computational efficiencies of the multi-grid algorithm, we compared the iteration times and time cost of the two methods, the results are shown in \ref{table1} where the time costs are the results on a notepad (2.8GHz CPU+12GB RAM).
\begin{table}[htbp]
	\caption{Comparison between the computational cost of the $ 1\times 1\times 1 $unit cell using the multigrid (MG) algorithm and the single-grid (SG) algorithm}\label{table1}
	\vspace{0.5em}\centering
	\begin{tabular}{cccccc}
		\toprule[1.5pt]
		\backslashbox{items}{$ h_{[0]} $/Bohr}&$ \dfrac{1}{10} $&$ \dfrac{1}{8} $& $ \dfrac{1}{6} $&$ \dfrac{1}{4} $&$ \dfrac{1}{2} $\\
		\midrule[1pt]
		The computing scale (number of grid points) &$ 199^3$ &$ 159^3$ & $ 119^3$ & $ 79^3$ & $ 39^3$\\
		The iteration times at level 0 of MG &12 & 11 & 20 & 8 & 7\\
		The iteration cycles of MG &3 & 3 & 4 & 2 & 2\\
		The iteration times of SG &36 & 29 & 23 & 18 & 12\\
		The time cost of MG/s &2040 &912 & 598 & 66 & 7\\
		The time cost of SG/s &5343 &1727 & 548 & 112 &8\\
		\bottomrule[1.5pt]
	\end{tabular}
\end{table}
Table \ref{table1} indicates that both the time costs of the MG and SG algorithms increases with decreasing $ h_{[0]} $, but the iteration cycles of the MG algorithm stably lies between $ 3\pm 1 $, making the increasing of time cost mainly caused by the increasing time cost of single iteration. Meanwhile, the number of iterations rapidly increase to three times the value of the MG algorithm when $ h_{[0]}=1/10 $Bohr. In addition, when $ h_{[0]} =0.5\ $Bohr, the MG algorithm did not converge at level 2 with $ \varepsilon=1\times 10^{-4} $, making the iteration at level 0 converges to incorrect results (not displayed). If setting $ \varepsilon=5\times 10^{-4} $, then we can obtain the results in Table \ref{table1}. Thus, for the cases where the SG algorithm is already very efficient (only costs 8s), the MG algorithm is not necessary, since the iteration process is more complicated, the MG algorithm not only cannot make good use of its advantage, but also will converge to incorrect results. Let the cell size expand to $ 5\times 5 \times 5 $, we still use the values of $ h_{[0]} $ above and further compared the computational cost of the two algorithms on a desktop computer (3.3GHz CPU+32GB RAM), and obtained the results in table \ref{table2}.
\begin{table}[htbp]
	\caption{Comparison between the computational cost of the $ 5\times 5 \times 5 $ unit cell using the multigrid (MG) algorithm and the single-grid (SG) algorithm}\label{table2}
	\vspace{0.5em}\centering
	\begin{tabular}{cccccc}
		\toprule[1.5pt]
		\backslashbox{items}{$ h_{[0]} $/Bohr}&$ \dfrac{1}{10} $ &$ \dfrac{1}{8} $& $ \dfrac{1}{6} $&$ \dfrac{1}{4} $&$ \dfrac{1}{2} $\\
		\midrule[1pt]
		The computing scale (number of grid points) &$ 523^3$ &$ 419^3$ & $ 315^3$& $ 211^3$& $ 107^3$ \\
		The iteration times at level 0 of MG &13 & 10 & 18& 11 & 11\\
		The iteration cycles of MG &3 & 2 & 4 &3&3 \\
		The iteration times of SG &28 & 23 & 18& 14& 14\\
		The time cost of MG/s &20081 &7112& 5138&791& 72\\
		The time cost of SG/s &38081 &13678 & 4261&674 & 51 \\
		\bottomrule[1.5pt]
	\end{tabular}
\end{table}
Table \ref{table2} indicates that both the computational costs of the MG and the SG algorithms increases rapidly with the decreasing mesh size, while the increasing rate of the former is much slower than the later. For this cell, the iteration cycles of the MG algorithm still stably lie between $ 3\pm 1 $. When $ h_{[0]} \leqslant \frac{1}{8}$Bohr, the computational efficiency of the MG algorithm is about 1.9 times as high as that of the SG algorithm. For large mesh sizes, since the electron density contains smaller ratio of the low-frequency components, the iteration number needed for the convergence of the SG algorithm is sufficiently small (no more than 18), thus, the MG algorithm cannot fully show its advantage and the time cost is slightly longer than that of the SG algorithm. 

Table \ref{C2tim} further compared the time cost of the proposed SG algorithm with the values in the literature.
\begin{table}[htbp]
	\caption{Comparison between the time cost (Unit: s) of the methods in the literature and that of the proposed method.}\label{C2tim}
	\vspace{0.5em}\centering
	\begin{tabular}{ccc}
		\toprule[1.5pt]
		cell size (the number of atoms) & methods in the literature  & the proposed SG algorithm\\
		\midrule[1pt]
		$ 5\times 5 \times 5 $\ (666)&259\cite{Ho2008},\ $ 3.6\times 10^{4} $\cite{Suryanarayana2014}&51 \\
		$ 10\times 10 \times 10 $\ (4631)&$ 4.2\times 10^3 $\cite{Mi2016} & 390 \\
		$ 29\times 29 \times 29 $\ (102690)&$ 1.224\times 10^6 $\cite{Suryanarayana2014},\ $ 1.152\times 10^5 $\cite{Shao2021} & $ 1.5596\times 10^4 $  \\
		\bottomrule[1.5pt]
	\end{tabular}
\end{table}
Table \ref{C2tim} indicates that the time costs of the proposed SG algorithm are lower than that of the literature by one to two orders of magnitude, indicating the superiority of the proposed SG algorithm in computational efficiencies.

\section{Conclusions}
This work developed a new solver for real-space OFDFT, including six algorithms: Algorithm \ref{uSG} is the single-grid algorithm, Algorithm \ref{LineSearch} is for proceeding line search in Algorithm \ref{uSG} to obtain the optimal step length in each iteration, Algorithm \ref{FgF} is for inputting Algorithm \ref{der} in order to provide information of the physical quantities during the line search, Algorithm \ref{WolfeCheck} is to judge whether terminate the line search process, Algorithm \ref{algMG} is the multi-grid algorithm based on the previous five algorithms. By directly solving the inner Poisson equation and proposing a new line search method for the outer iteration, the new solver enhanced the computational efficiencies by one to two orders of magnitude compared with that of the literature. The multi-grid algorithm can further improve the computational efficiencies by one time (if the memory is sufficient).

\bibliography{reference}
\end{document}